\documentclass[fleqn,usenatbib]{mnras}

\usepackage{newtxtext,newtxmath}
\usepackage{systeme}

\usepackage[T1]{fontenc}
\usepackage{ae,aecompl}
\usepackage{relsize}
\usepackage[english]{babel}
\usepackage[utf8x]{inputenc}
\usepackage[T1]{fontenc}
\usepackage{natbib}
\usepackage{subcaption}
\usepackage{verbatim}
\usepackage{siunitx}
\usepackage{booktabs}
\usepackage{graphicx}	
\usepackage{amsmath}	
\usepackage{breqn}

\newcommand{\bl}{\color{blue}}

\title[The sizes of compact MGs]{The evolution of compact massive quiescent and starforming galaxies derived from the $R_e-R_h$ and $M_{\rm star}-M_h$ relations}
\author[L. Zanisi ]{
L. Zanisi$^{1,2}$ \thanks{E-mail: L.Zanisi@soton.ac.uk},
F. Shankar$^{1}$ , H. Fu$^{1}$, A. Rodriguez-Puebla$^{3}$, V. Avila-Reese$^{3}$,  A. Faisst$^{4}$,  E. Daddi$^{5}$, \newauthor L. Boco$^{6}$,  A. Lapi$^{6,7}$, M. Giavalisco$^{8}$, P. Saracco$^{9}$, F. Buitrago$^{10,11}$, M. Huertas-Company$^{12,13}$,\newauthor
 A. Puglisi$^{14}$, A. Dekel$^{15}$
\\
$^{1}$ Department of Physics and Astronomy, University of Southampton, Highfield, SO17 1BJ, UK\\
$^{2}$ DISCnet centre for doctoral training, University of Southampton, Highfield, SO17 1BJ, UK\\
$^{3}$ Instituto de Astronom\'ia, Universidad Nacional Aut\'onoma de M\'exico, A. P. 70-264, 04510, M\'exico, D.F., M\'exico\\
$^{4}$ PAC, California Institute of Technology 1200 E California Boulevard, Pasadena, CA 91125, USA\\
$^{5}$ CEA, IRFU, DAp, AIM, Universit\`{e} Paris-Saclay, Universit\`{e} Paris Diderot, Sorbonne Paris Cit\`{e}, CNRS, F-91191 Gif-sur-Yvette,France\\
$^{6}$ SISSA, Via Bonomea 265, I-34136 Trieste, Italy\\
$^{7}$ IFPU - Institute for fundamental physics of the Universe, Via Beirut 2, 34014 Trieste, Italy\\
$^{8}$ Department of Astronomy, University of Massachusetts Amherst, 710 N. Pleasant St., Amherst, MA 01003, USA\\
$^{9}$ INAF - Osservatorio Astronomico di Brera, via Brera 28, 20121 Milano, Italy\\
$^{10}$ Departamento de F\'{i}sica Te\'{o}rica, At\'{o}mica y \'{O}ptica, Universidad de Valladolid, 47011 Valladolid, Spain\\
$^{11}$ Instituto de Astrof\'{i}sica e Ci\^{e}ncias do Espa\c{c}o, Universidade de Lisboa, OAL, Tapada da Ajuda, PT1349-018 Lisbon, Portugal\\
$^{12}$ Instituto de Astrof\'isica de Canarias (IAC); Departamento de Astrof\'isica, Universidad de La Laguna (ULL), E-38200, La Laguna, Spain\\
$^{13}$ LERMA, Observatoire de Paris, CNRS, PSL, Universit\'e Paris Diderot, France\\
$^{14}$ Center for Extragalactic Astronomy, Durham University, South Road, Durham DH1 3LE, United Kingdom\\
$^{15}$  Racah Institute of Physics, The Hebrew University, Jerusalem 91904, Israel
}

\date{Accepted for publication in MNRAS}

\begin{document}
\label{firstpage}
\pagerange{\pageref{firstpage}--\pageref{lastpage}}
\maketitle



\begin{abstract}
The mean size ( effective radius $R_e$) of Massive Galaxies  (MGs, $M_{\rm star}>10^{11.2}M_\odot$) is observed to increase steadily with cosmic time. It is still unclear whether this trend originates from the size growth of individual galaxies (via, e.g., mergers and/or AGN feedback) or from the inclusion of larger galaxies entering the selection at later epochs (progenitor bias). We here build a data-driven, flexible theoretical framework to probe the structural evolution of MGs. We assign galaxies to dark matter haloes via stellar mass-halo mass (SMHM) relations with varying high-mass slopes and scatters $\sigma_{\rm SMHM}$ in stellar mass at fixed halo mass, and assign sizes to galaxies using an empirically-motivated, constant and linear relationship between $R_e$ and the host dark matter halo radius $R_h$. We find that: 1) the fast mean size growth of MGs is well reproduced independently of the shape of the input SMHM relation; 2) the numbers of compact MGs grow steadily until $z\gtrsim2$ and fall off at lower redshifts, suggesting a lesser role of progenitor bias at later epochs; 3) a time-independent scatter $\sigma_{\rm SMHM}$ is consistent with a scenario in which compact starforming MGs transition into quiescent MGs in a few $10^8$yr with a negligible structural evolution during the compact phase, while a scatter increasing at high redshift implies significant size growth during the starforming phase. A robust measurement of the size function of MGs at high redshift can set strong constraints on the scatter of the SMHM relation and, by extension, on models of galaxy evolution.
\end{abstract}


\section{Introduction}
\label{sec:introduction} 
There is now substantial evidence that galaxies of a given stellar mass are smaller at higher redshift than in the local Universe (e.g., \citealt{Daddi+05}, \citealt{Trujillo+07},\citealt{Buitrago+08}, \citealt{vandokkum+10}, \citealt{Cassata+11}, \citealt{Cimatti+12},\citealt{Newman+12}, \citealt{Huertas-company+13_sizeEv}, \citealt{vandokkum+15}, \citealt{Kawamata+15}, \citealt{Shibuya+15}). The size evolution of the galaxy population in a given stellar mass bin is well fitted by a relation of the type
\begin{equation}
\label{eq:Rez}
R_e(z) \propto (1+z)^{-\alpha}
\end{equation}
where $R_e$ is defined as the radius that encloses half of the galaxy light (see, e.g., \citealt{vanderwel+2014} for a different fitting function). It is found that in general starforming galaxies follow shallower trends (lower values of $\alpha$) than quiescent galaxies  (e.g., \citealt{vanderwel+2014}). Notably, the size growth rate of starforming galaxies increases with stellar mass, becoming comparable to that of quiescent galaxies (with $\alpha\sim 1$) for $M_{\rm star}>10^{11.2} M_\odot$ (\citealt{Faisst+17}, \citealt{Mowla+18}). Moreover, in this mass regime, as pointed out in several studies (e.g., \citealt{Bernardi+11b,Bernardi+11a,Cappellari2016_review}), the behaviour of the scaling relations and the stellar kinematics differ from that of less massive galaxies. Thus, the mass scale $M_{\rm star}\approx10^{11.2}M_\odot$ is critical to understanding galaxy evolution. In this paper we focus on the structural evolution of galaxies in this high mass regime, which we simply label in what follows as ``Massive Galaxies (MGs)''.

There is no consensus yet as to why MGs were a factor of 3 to 5 smaller in the past. Minor dry mergers have been invoked as an efficient channel to promote substantial size increase with relatively modest change in stellar mass (e.g.,  \citealt{Naab+09},\citealt{Oser+10},\citealt{Shankar+13}, \citealt{vandokkum+15}) to accomodate for the limited evolution in the high-mass end of the stellar mass function (SMF) since $z\sim1.5$ (e.g.,\citealt{Andreon2013},\citealt{Muzzin+13}, \citealt{McDermid+15}, \citealt{Kawinwanichakij+20}). However,  the rate of minor dry mergers may not be  sufficient by themselves to account for the entire size evolution of MGs through cosmic time \citep{Newman+12,Nipoti+09,Nipoti+12}. 
More generally, the exact contribution of dry mergers to the mass assembly of massive galaxies
is still a matter of intense debate among both theoretical studies (e.g., \citealt{DeLucia&Blaizot2007,Hopkins+10_mergerRates, Wilman+13_morpho_hierarchical_SAM,Rodriguez-Gomez+15,Qu+18_massAssembly_EAGLE, Tacchella+19,OLeary+20_mergerRates, Grylls+20_pairfract}) and observational works (e.g., \citealt{Man+16_mergerRates,Mundy+17,Mantha+18,Duncan+19_mergerRates}).

It has often been debated in the literature whether the size evolution of galaxies of a given stellar mass stems from the size growth of individual galaxies or it is a consequence of a ``population effect'' where newly formed, larger galaxies enter the mass selection at later epochs thus increasing the mean size distribution (e.g., \citealt{Carollo+13}, \citealt{Shankar+15}, \citealt{Gargiulo+17}). This ``progenitor bias" effect \citep{vanDokkum&Franx1996} has been usually invoked to explain the size evolution of passive galaxies with $M_{\rm star}<10^{11}M_\odot$  (e.g. \citealt{Faisst+17, Fagioli+16}). Most studies agree on the lesser role of progenitor bias in the size evolution of MGs at $z\lesssim 1$, in favour of a more predominant role of (dry) mergers in increasing the sizes of individual galaxies (e.g.,\citealt{Saglia+10}, \citealt{Carollo+13}, \citealt{vanderwel+2014}, \citealt{Fagioli+16}, \citealt{Faisst+17}, but see also \citealt{Gargiulo+17}). In particular, the disappearance of compact (e.g., \citealt{Cassata+11,Barro+13}) galaxies as the Universe ages is interpreted as a sign that they must have grown in size individually \citep{vanderwel+2014} while a constant abundance of compact galaxies implies that progenitor bias dominates the size growth \citep{Saracco+10,Gargiulo+16,Gargiulo+17}. In this respect, the full distribution of galaxy sizes at fixed stellar mass, i.e., the size function $\phi(R_e|M_{star})$, is an invaluable tool to disentangle galaxy evolution scenarios, providing simultaneous information on the mean size $R_e$ and the number density of compact galaxies (e.g., \citealt{Shankar+10_sizefunct, Carollo+13}, {\bl{Z20}}). 

In addition to mergers and progenitor bias, Active Galactic Nuclei (AGN) feedback during the  compact starforming stages of the evolution of MGs (both in a submm-FIR phase, e.g. \citealt{Barro+16}, and an optical ``blue nugget'' phase, e.g. \citealt{Martig+09,Damjanov+11,Barro+13,Fang+13,Zolotov+15,Tacchella+16}, which are potentially linked in an evolutionary sequence, e.g. \citealt{GomezGuijarro+19_compacts_old_starbursts,Puglisi+21}) may also contribute to both size growth and quenching \citep{Fan+08,Fan+10,Kocevski+17,Lapi+18_ETGs, Vandervlugt+19}, and the relative evolution of compact starforming and quiescent galaxies can provide tight constraints on these processes, as we will further discuss below.

Due to their flexibility and relatively lower number of free parameters, semi-empirical models have become a popular route to study the mass assembly, star formation, and merger histories of galaxies (e.g., \citealt{Conroy&Wechsler2009,Hopkins+10_whichMergerMatters,Hopkins+10_mergerRates,Behroozi+13,Moster+13,Shankar+13,Gu+16_scatterSMHM,Matthee+17,Tinker+17_testingTheoriesWithScatter,Rodriguez-Puebla+17,Lapi+18_ETGs, Chen+20_Faber_semiempirical,Grylls+20_satellites,Grylls+20_pairfract,OLeary+21}). The main ingredient in semi-empirical models is the input stellar mass-halo mass (SMHM) relation, which is extracted from the cumulative equivalence between the number densities of the stellar mass and (sub)halo mass functions \citep[e.g.,][]{Vale&Ostriker2006,Shankar+06,Guo+11,Dutton+10,Firmani&AvilaReese2010,Leauthaud+12,Rodriguez-puebla+13, Mandelbaum+16,Pillepich+18,Behroozi+18, Moster+18, Kravtsov+18,Erfanianfar+19}.   However, the systematic uncertainties in the input data, most notably in the stellar mass function\footnote{{\label{footnote:intro}Ultimately, the reasons for these discrepancies are thought to originate from the way stellar masses are estimated. The Initial Mass Function, dust attenuation curve, stellar population synthesis models and assumed star formation histories, the inclusion of intra-cluster light and even the photometry choice and background subtraction algorithms all contribute to the various determinations of the stellar mass functions found in the literature (see, e.g., \citealt{Bernardi+10,Bernardi+13,Bernardi+16,Bernardi+17_highmassend, Kravtsov+18,Leja+20, Lower+20,Guarnieri+20, Kawinwanichakij+20} amongst many others), which result in different estimates of the SMHM relation (e.g., \citealt{Shankar+17_bulgehalo})}}, have yielded discrepant results along the years. For example, some groups \citep[e.g.,][]{Moster+13,Behroozi+13,Rodriguez-Puebla+17} proposed a shallower high-mass slope in the SMHM relation, while others have argued in favour of a steeper slope \citep[e.g.,][]{Shankar+14_SMHM,Tinker+17_SMHM_massivegal,Kravtsov+18,Grylls+19}. In turn, the high-mass end of the SMHM relation plays a crucial role in, e.g., the number of galaxy pairs, and thus on galaxy merger rates \citep{Grylls+20_pairfract,OLeary+21}.

Some works also highlighted a further correlation between the size of a galaxy, $R_e$, and that of its host dark matter halo, $R_h$ (e.g. \citealt{Fall1983, MoMaoWhite98,Kravtsov2013,Huang+17,Desmond2017_concentration,Hearin+19, Desmond+18_SPARC, Somerville+18, Lapi+18_disks,Mowla+19_SMHM_R80, Zanisi+20}, but see also \citealt{Desmond+17_EAGLE} ). When coupled together in a semi-empirical model, the SMHM and $R_e-R_h$ relations become powerful tools to simultaneously probe the mass and structural evolution of galaxies in a full cosmological context. For instance, \citealt{Somerville+18} found a non-trivial redshift and stellar mass dependence of the mean $R_e-R_h$ relation at $M_{\rm star}\lesssim10^{11.2}M_\odot.$ 
\citet{Zanisi+20} (hereafter {\color{blue} Z20}), found that, in the MGs regime and at $z\sim0.1$, the $R_e-R_h$ relation must be very tight with a total scatter (inclusive of observational statistical uncertainties) of only $\approx0.1$ dex\footnote{This was necessary to reproduce the full size distribution of early- and late-type MGs in the local Universe as measured in the Sloan digital Sky Survey (SDSS).}
\citet{Stringer+14} (hereafter {\bl{S14}}) combined the \citet{Moster+13} SMHM relation and a constant $R_e-R_h$ relation to build a semi-empirical model which proved capable to reproduce the size evolution of MGs in the COSMOS field \citep{Huertas-company+13_sizeEv}. {\bl{S14}} attributed the mean size growth of the population of MGs at $z\lesssim 2$ to a cosmological effect for which (i) the size of the host dark matter haloes of a given mass become larger as the Universe expands and its density decreases and (ii) MGs of the same mass form in more massive, extended dark matter haloes at lower redshift. While these results are encouraging, the effects of assuming a different SMHM relation in the framework outlined by {\bl{S14}} has remained relatively unexplored. The only notable exception is the work by \citet{Mowla+19_SMHM_R80}, where however only the shape of the SMHM was considered and not its scatter, $\sigma_{\rm SMHM}$. The parameter $\sigma_{\rm SMHM}$ is notoriously degenerate with the high-mass slope of the SMHM relation in retrieving the number density of massive galaxies \citep{Shankar+14_SMHM, Wechsler&Tinker2018}. As we will show, instead, $\sigma_{\rm SMHM}$ has a seizable effect on the number density of the population of compact galaxies, and therefore it is a significant novelty that we include in the semi-empirical framework set out in {\bl{S14}}.

\begin{figure*}

    \centering
    \includegraphics[width=\textwidth]{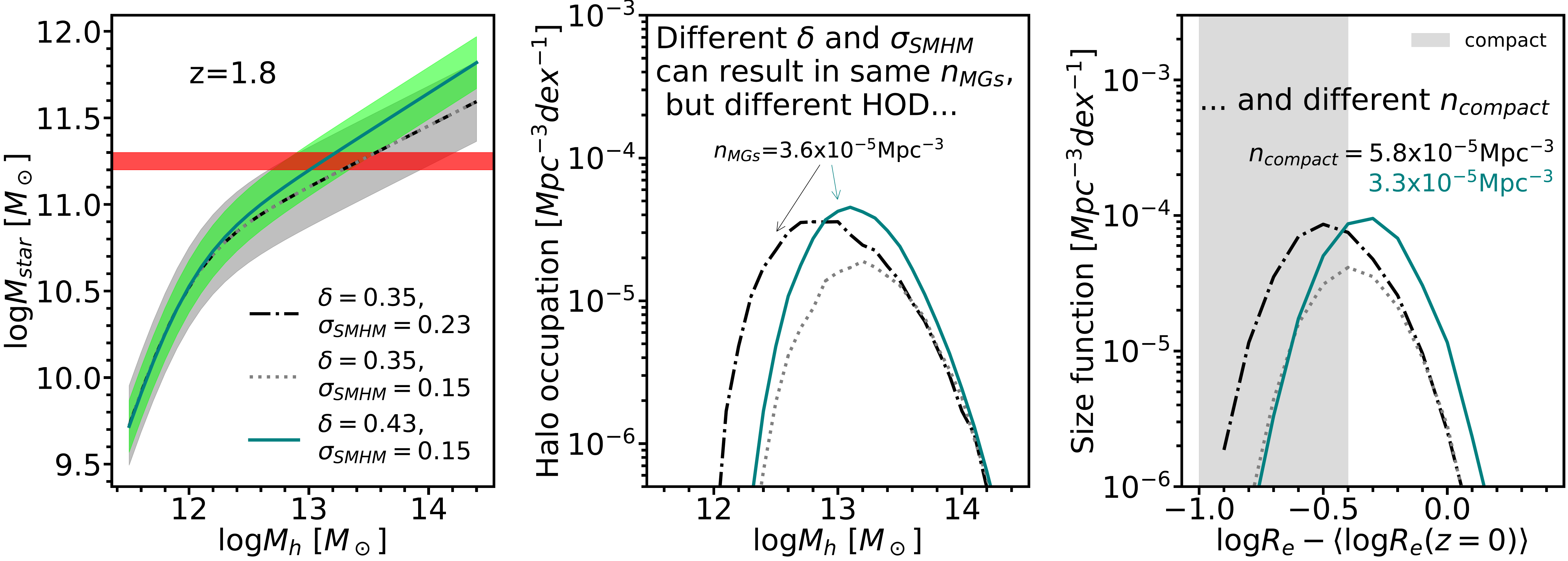}
    \caption{ The effect of different SMHM relations on the halo occupation distribution (HOD) of MGs and the size functions implied by a linear $R_e-R_h$ relation (eq. \ref{eq:K13model}). We show that SMHM relations with different high-mass slope $\delta$ and scatter $\sigma_{\rm SMHM}$ (shown as shaded areas in the left panel) can produce the same number density for MGs (the mass threshold for MGs is shown as a red horizontal line). However, the halo occupation distribution for the two models is remarkably different (central panel). This translates in very different size functions. In particular, the number density of compact galaxies differs by almost a factor of two (we use the \citealt{Cassata+13} definition of compactness, i.e. 0.4 dex below the z=0 mean size, against which we calibrated the two models following Appendix \ref{app:sloan} ). The model indicated with dashed gray lines is shown to help appreciate the effect of a lower $\sigma_{SMHM}$, at fixed $\delta$, on the halo occupation distribution and on the size function (compare to the dot-dashed model). } Although we show results only at a given redshift as an example, the same arguments apply at any epoch for some choices of $\delta$ and $\sigma_{\rm SMHM}$.
    \label{fig:cartoon_hod_introduction}
\end{figure*}

In this paper we put forward a phenomenological, transparent methodology to probe the size evolution of massive galaxies that expands on the works by {\bl{S14}}, \citealt{Somerville+18} and \citet{Mowla+19_SMHM_R80}. Following the approach of \citet{Grylls+20_pairfract}, we build mock catalogs of galaxies in dark matter haloes using data-driven toy models where SMHM relations with different shapes and dispersion are coupled with { linear $R_e-R_h$ relations characterized by different normalizations}. The main objective of this paper is to probe the impact of varying the input SMHM relation and its dispersion $\sigma_{\rm SMHM}$ on: 1) the mean size evolution of MGs, 2) the full size function of MGs across cosmic time, and 3) the number density of compact MGs. The latter point is particularly original and powerful as the time dependence of the number density of compact MGs is closely linked to progenitor bias: less compact galaxies at fixed stellar mass would be observed at later epochs if they grow in size via, e.g., mergers. 
We will show, in particular, that the scatter in the SMHM relation plays a major role in setting the number density of compact MGs, allowing to break the degeneracy between the scatter and the high-mass slope of the SMHM (see, e.g., Figure \ref{fig:cartoon_hod_introduction}). 
Our present work lays out an effective strategy to unveil the evolutionary pathways of MGs by exploiting the increased statistics of MGs that will become available from future observations.
Data for MGs are in fact at present quite sparse and uncertain at $z\gtrsim 1$ (e.g., \citealt{Kawinwanichakij+20}), and effective radii have been measured for only a handful of MGs at $z\gtrsim 2$ (e.g., \citealt{Kubo+17,Patel+17, Faisst+17, Mowla+18,Stockmann+20,Lustig+21}).

The outline of the paper is as follows. In Section \ref{sec:methods} we provide the backbone of our framework. In Section \ref{sec:core} we explore the role of the scatter in the SMHM relation and we present toy models inspired to these findings in Section \ref{sec:systematic_analysis}.  In Sections \ref{sec:hod}, \ref{sect:size_evolution} and \ref{sec:fquench_compact} we show how the different SMHM implied by the toy models results in a range of possible determinations of the size distributions of MGs. We give an interpretation of the evolutionary pathways of MGs implied by the different toy models in Section \ref{sec:progbias}, and we show how our framework can be used to constrain the shape and scatter of the SMHM relation in Section \ref{sec:sizes_constraints_SMHM}. In Sections \ref{sec:covariance} we discuss the possibility that the SMHM and $R_e-R_h$ relations are correlated, and in Section \ref{sec:concentration} we discuss an extension of the model of the $R_e-R_h$ connection based on halo concentration, and discuss its limitations and strengths. Finally, we draw our conclusions in Section \ref{sec:conclusions}. Further material is available in the Appendices.

\begin{figure*}
    \centering
    \includegraphics[height=\dimexpr \textheight - 10\baselineskip\relax]{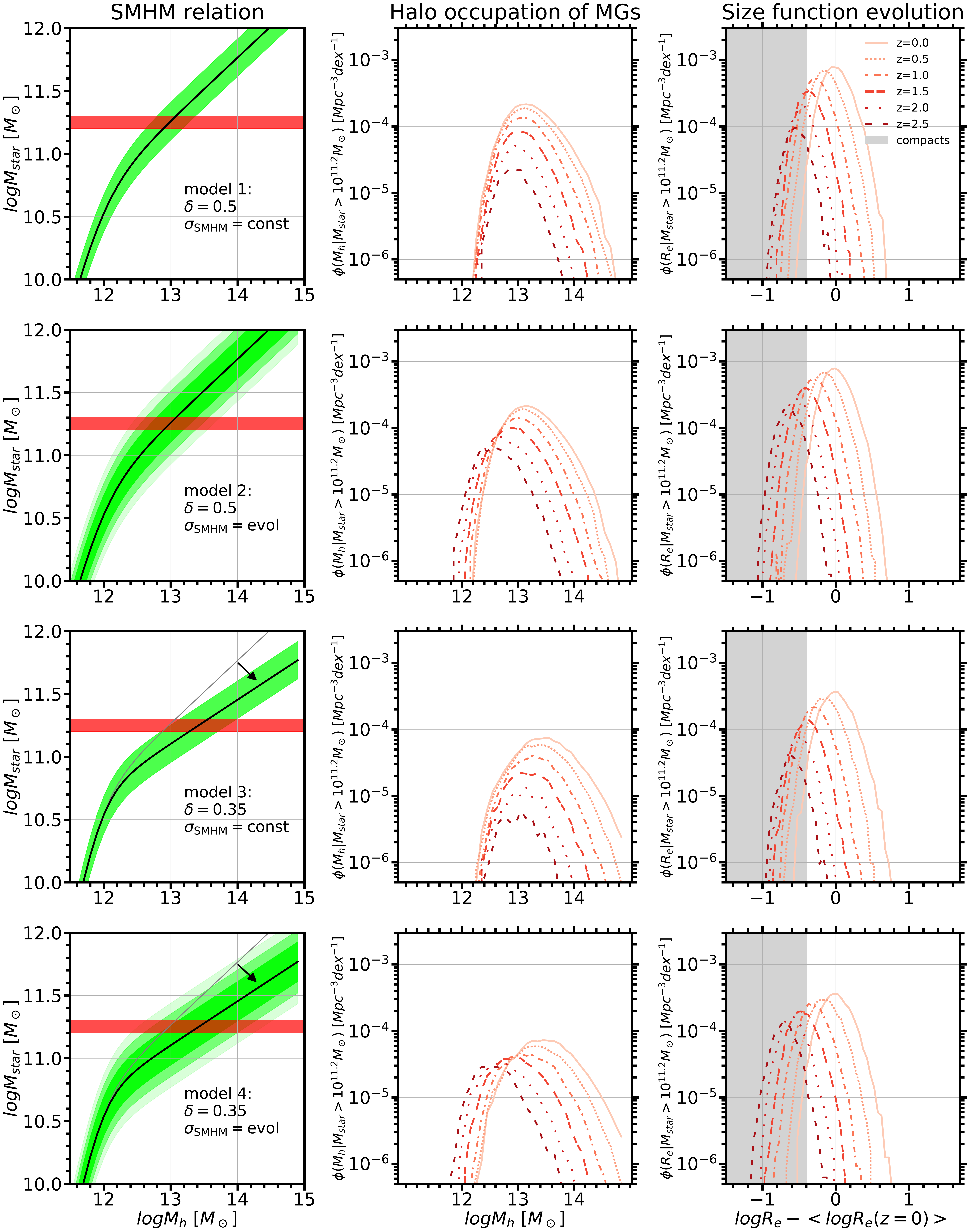}
    \caption{ \emph{Left column:} The SMHM relation of the four models outlined at the beginning of Section \ref{sec:systematic_analysis}. The red line indicates the stellar mass selection for MGs. The green shaded regions indicate the scatter of the SMHM, which increases at higher redshift for models 2 and 4. The gray line in the center-bottom and bottom panels indicates the SMHM for Models 1 and 2 as a reference. \emph{Central and Right column:} The redshift evolution of the halo occupation distribution $\phi(M_h|M_{\rm star}>10^{11.2}M_\odot)$ and the implied size functions $\phi(R_e|M_{\rm star}>10^{11.2}M_\odot)$  of MGs for the four models. We display results for z=0,0.5,1,1.5,2,2.5,3. Darker colours indicate higher redshift. The gray band in the right column shows the \citet{Cassata+13} definition for compact galaxies. It can be seen that the increasing $\sigma_{\rm SMHM}$ of Models 2 and 4 results in broader distributions, which have a median lower $M_h$ and normalised $R_e$ compared to Models 1 and 3, where $\sigma_{\rm SMHM}=0.15$ dex at all times. An evolving $\sigma_{\rm SMHM}$ also results in a higher number density of MGs at earlier times.  Contrariwise, the flatter high-mass-end slope of the SMHM in Model 3 results in overall fewer MGs and slightly larger median halo masses compared to Model 1. }
    \label{fig:cartoon_hod}
\end{figure*}

\section{Methods}
\label{sec:methods}

\subsection{The galaxy-halo connection}
\label{sec:galaxyhalo}
 To study the size distribution and evolution of MGs ($M_{\rm star}>10^{11.2} M_\odot$), at any redshift of interest we build a catalogue of dark matter haloes with mass $M_h$ and size $R_h$, to which we assign a stellar mass $M_{star}$ (via an input SMHM relation) and an effective radius $R_e$ (via a  $R_e$-$R_h$ relation).
The detailed modelling approach that we use here closely follows the one outlined in \citet{Zanisi+20}, which we briefly summarise below for convenience\footnote{We make extensive use of the \texttt{COLOSSUS} Python package \citep{DiemerColossus}.}:

\begin{enumerate}
\item We use the \citet{Despali+16} halo mass function to obtain large catalogues of dark matter haloes at all the redshifts of interest. Note that the \citet{Despali+16} halo mass function is defined for central galaxies only, as it does not include subhaloes (e.g. \citealt{vandenbosch*14}). Thus, all MGs in this study are modelled as central galaxies.\footnote{In our model all MGs are considered central galaxies, since satellites are negligible in this extreme mass range (e.g., \citealt{Peng+10}). Using the Statistical SemiEmpirical Model STEEL \citep{Grylls+19}, we find that the satellite contribution to MGs is less than 20\% at z$\sim 0.1$, and it declines steeply at earlier times. The subdominant population of satellite MGs is studied in a companion paper (Zanisi et al. 2020).}.
In this paper we model dark matter haloes with a \citet{NavarroFrenkWhite95} density profile with scale radius $R_s$,
\begin{equation}
\rho(r) \propto \frac{1}{\frac{r}{R_s}\bigl[ 1+\frac{r}{R_s} \bigr]^2},
\end{equation}
and with $R_h=cR_s$ defining the \emph{concentration} parameter $c$. $R_h$ is the dark matter halo radius,
\begin{equation}
R_{h}=\Bigl( \frac{3M_{h}}{4\pi \cdot \Delta\rho_{\Delta}} \Bigr)^{\frac{1}{3}}
\label{eq:Rhalo}
\end{equation}
 where $\Delta$ is the virial overdensity with respect to the cosmological critical density \citep{Bryan&Norman98}. Both $\Delta$ and $\rho_{\Delta}$ are decreasing functions of cosmic time (e.g, \citealt{Mo+10}). Thus, a dark matter halo of a given mass has a smaller size at higher redshift, owing to the higher density of the Universe.
\item We model the link between galaxies and dark matter via the stellar mass-halo mass relation (SMHM). The mean SMHM$\equiv M_{\rm star}(M_{\rm h})$ is a monotonically increasing function of halo mass. We include a lognormal scatter $\sigma_{\rm SMHM}$ at fixed halo mass that takes into account both the intrinsic dispersion $\sigma_{\rm intr}$ in the relation and the uncertainty in stellar mass estimates $\sigma_*$ (i.e., $\sigma_{\rm SMHM}^2 = \sigma_{\rm intr}^2 + \sigma_{*}^2$, see \citealt{Rodriguez-Puebla+17}, \citealt{Behroozi+13}, \citealt{Shankar+14_SMHM}, \citealt{Tinker+17_SMHM_massivegal}). In the next Section we will present ``toy'' SMHM relations, which vary in both shape and dispersion, to probe their impact on our galaxy mocks and on their size distributions at different epochs. In particular, we will focus on the slope above the knee of the SMHM relation, $\delta$ (see Figure \ref{fig:cartoon_hod_introduction}), which is the parameter in the SMHM relation controlling the number density of MGs at a given dispersion. The precise value of $\delta$, or better of the underlying abundances of MGs in the local and high redshift Universe, still suffer from substantial systematic uncertainties and that are hotly debated in the literature (see Section \ref{sec:introduction}).

\item We assign a half light radius $R_e$ to each galaxy according to the ansatz:
\begin{equation}
R_e = A_K R_h,
\label{eq:K13model}
\end{equation}
which is based on the empirical findings by \citet{Kravtsov2013}, and that we call the \emph{K13 model}. 
Here $A_K$ is the normalization which in principle may vary with halo mass, galaxy stellar mass and/or star formation activity (e.g. \citealt{Huang+17,Somerville+18}. {\bl{Z20}}, \ref{app:sloan}). We add to eq. \ref{eq:K13model} an intrinsic log-normal scatter $\sigma_K$, which, as $A_K$, is a free parameter. We stress that, effectively, $R_e$ is a function of $M_h$, since there is a direct proportionality between halo mass and halo radius (see eq. \ref{eq:Rhalo}). While in the remainder of the paper we will mostly comment on the K13 model, in Section \ref{sec:concentration} we discuss another model of galaxy sizes in which the relation between virial radius and galaxy size is also mediated by the halo concentration (e.g., \citealt{Desmond2017_concentration, Jiang+19, Zanisi+20}). Moreover, other definitions of galaxy sizes, such as $R_{80}$ \citep{Miller+19_R80} or $R_1$ \citep{Trujillo+20}, have been proposed to correlate to $R_h$ equally well or even better than effective radius. We will discuss these models in Appendix \ref{sec:other_sizes}.
\end{enumerate}

\subsection{Quenching}
\label{sec:fquench}
 To provide a fair comparison to observations, which have so far always distinguished between starforming and quiescent MGs \citep[e.g.,][]{Mowla+18}, we need to include a recipe for quiescence in our galaxy mocks.
To this purpose, following the empirical calibration of \citet{Rodriguez-Puebla+15} at $z\sim 0.1$, we assume that the probability of a galaxy being quenched in a dark matter halo of mass $M_h$ is { given by the fraction}
\begin{equation}
f_{Quench} (M_h)= \frac{1}{b_0 + [\mathcal{M}_0 \times10^{12}/M_h (M_{\odot})]}
\label{eq:fQevol}
\end{equation}
with $b_0 \sim 1$ and $\mathcal{M}_0 \sim 0.68$ at $z \sim 0.1$. $f_{Quench}$ is a monotonically increasing function of halo mass, with a characteristic mass scale $\mathcal{M}_0$ above (below) which more (less) than 50\% of galaxies are quiescent (starforming).

The fraction of quenched MGs is observed to evolve with redshift (e.g. \citealt{Huertas-Company+16, Mowla+18}). While it is beyond the scope of this work to set specific constraints on the physical processes that drive quenching (see \citealt{Somerville&Dave2015} for a review), we note that quenching is thought to be more likely to occur in more massive haloes at higher redshift (e.g., see the empirical models by \citet{Rodriguez-Puebla+17} and \citet{Behroozi+18}). In our model this is achieved by replacing $\mathcal{M}_0$ with 
\begin{equation}
\label{eq:betaz}
\mathcal{M} (z) = \mathcal{M}_0 + (1+z)^\mu,
\end{equation}
where $\mu>0$ is a free parameter, that regulates the increase in characteristic quenching halo mass in the younger Universe. Figure \ref{fig:fquench_Mh_z} shows examples of the evolution in $f_{Quench}$ for $\mu=1,3,5$.  We note that quiescence is defined in the literature according to different methods (e.g. 1 $\sigma$ below the main sequence, different cuts in the color-color planes, a hard cut in specific star formation rate, see e.g., \citealt{Donnari+19}) which can lead to different results \citep{Sherman+20}. Therefore, the value of $\mu$ will depend on the method assumed. For this reason, in the following we simply show different values of $\mu$, which we will adapt to the specific method used once the comparison data is fixed.

Quiescent and starforming MGs of similar mass appear to grow in size at the same rate with redshift, with quiescent galaxies being systematically smaller at all times (\citealt{Mowla+18}). We assume that the two populations live on two separate K13 relations. The normalizations $A_{K,SF}$ and $A_{K,Q}$, for starforming and quenched MGs respectively, are calibrated at $z\sim 0.1$ following Appendix \ref{app:sloan}. Following {\bl{Z20}} we also assume that the scatters in the two K13 relations, $\sigma_{K,SF}$ and $\sigma_{K,Q}$, are equal to 0.1 dex. In the remainder of this paper, we assume that this value of $\sigma_K$ holds at all times.

\subsection{Target observables}
\label{sec:target}
Using the methodology outlined above, we will present the results of some toy models (described in Section \ref{sec:systematic_analysis}) for the following observables:
\begin{itemize}
\item the evolution of the galaxy size distribution of MGs (i.e. the size function $\phi(R_e,z|M_{\rm star}>10^{11.2}M_\odot)$) and its integral, the number density of MGs
    \begin{equation}
        n_{MGs}(z) = \int_{-\infty}^{\infty}\phi(R_e,z|M_{\rm star}>10^{11.2}M_\odot)dlogR_e;
    \end{equation}
    \item the mean size of the population of MGs as a function of redshift, $\langle R_e(z) \rangle$;
    \item the evolution in the number density of compact MGs $n_{compact}(z)$. A range of definitions of compactness have been proposed in the literature (e.g., \citealt{Saracco+10,Fang+13,Carollo+13,Barro+13,vanderwel+2014,Damjanov+15_numbers_compacts, vandokkum+15,Barro+17,Charbonnier+17_compacts, Tacchella+17,Buitrago+18,Tortora+18,Luo+20} amongst many others). Here we define galaxies as compact systems if their size is 0.4 dex below the $z\sim 0$ $R_e-M_{\rm star}$ relation of quenched galaxies \citep{Cassata+11, Cassata+13},
    \begin{equation}
    \label{eq:ncompact}
    n_{compact}(z) = \int_{-\infty}^{-0.4}\phi(R_e/R_e(z=0),z)dlog(R_e/ R_e(z=0)).
\end{equation}
    In particular, we focus on compact quenched MGs (CQMGs) and compact starforming MGs (CSFMGs). 
\end{itemize}

{ Figure \ref{fig:supplementary_compacts} shows that adopting other definitions of compactness based on the effective radius yields qualitatively similar results to the \citet{Cassata+11} definition.} Other popular definitions of compactness based on, e.g., the stellar mass density in the central kiloparsec, would require information on the light/mass profile of galaxies (e.g., the S\'ersic index), which we are not including here. This requires further modelling which we defer to future work.

\section{Results}
\label{sec:results}

\subsection{At the core of the model}
\label{sec:core}
The methodology outlined in Section \ref{sec:methods} makes use of only two ingredients: (i) the K13 relation (eq. \ref{eq:K13model}) and (ii) the SMHM relation (most notably the high-mass slope $\delta$ and the scatter $\sigma_{\rm SMHM}$). 

Figure \ref{fig:cartoon_hod_introduction} shows that two SMHM relation with different high-mass slope $\delta$ and scatter $\sigma_{\rm SMHM}$ are able to produce the same number density for MGs. The degeneracy between $\delta$ and $\sigma_{\rm SMHM}$ in producing the same abundances of massive galaxies was already identified in previous studies (e.g., \citealt{Behroozi+10, Shankar+14_SMHM}). What we emphasize here, for the first time to the best of our knowledge, is that the corresponding halo mass distributions (middle panel, see also \citealt{Shankar+14_SMHM}), and thus the implied size functions computed via the linear $R_e-R_h$ relation (right panel), remain however significantly distinct, especially below the peaks of the distributions.
The larger abundances of compact MGs is mostly driven by a larger scatter in the input SMHM relation,  as can be inferred by comparing black dot-dashed and gray dashed lines in Figure \ref{fig:cartoon_hod_introduction}. Thus, the abundance of compact galaxies represents a valuable observable to break the degeneracy between $\delta$ and $\sigma_{\rm SMHM}$, allowing to set constraints on the degree of progenitor bias and ultimately to discriminate between different models of galaxy formation.

\subsection{Toy models}
\label{sec:systematic_analysis}

Motivated by the discussion above, we devise four toy models to show the effect of varying $\delta$ and $\sigma_{\rm SMHM}$ on our target observables (Section \ref{sec:target}):
\begin{itemize}
    \item Model 1: $\delta=0.5$ (steep slope), $\sigma_{\rm SMHM}=0.15$ dex at all redshifts;
    \vspace{0.2em}
    \item Model 2: $\delta=0.5$ (steep slope),
    $\sigma_{\rm SMHM}=\sqrt{(0.1 z)^2+0.15^2}$;
    \vspace{0.2em}
    \item Model 3: $\delta=0.35$ (shallow slope), $\sigma_{\rm SMHM}=0.15$ dex at all redshifts;
    \vspace{0.2em}
    \item Model 4: $\delta=0.35$ (shallow slope), $\sigma_{\rm SMHM}=\sqrt{(0.1z)^2+0.15^2}$.
\end{itemize}

 The slope of Model 1 (Model 2) is inspired to the \citet{Grylls+20_satellites} 'PyMorph' ('cmodel') SMHM relation, which was obtained by fitting the \citet{Bernardi+17_highmassend} 'PyMorph' ('cmodel') stellar mass function (SMF) at $z\sim 0.1$ and the \citet{Davidzon+17} SMFs at $z\gtrsim 0.2$ where their masses have been corrected by 0.15 dex to bring the two studies in agreement\footnote{This was done only for the 'PyMorph' SMF. The factor of 0.15 dex takes into account the difference in $M/L$ used in the two studies.} (see also \citealt{Bernardi+16}).\footnote{We also shift by -0.1 dex the knee of the SMF resulting from the \citet{Grylls+20_satellites} SMHM to better match the SDSS SMF.}

Although some authors point to distinct SMHM relations for quiescent and starforming galaxies (e.g., \citealt{Rodriguez-Puebla+15, Moster+18, Behroozi+18, Posit&Fall2021}), the relative content of stars in starforming and quiescent galaxies at fixed halo mass is still highly debated \citep[e.g.,][]{Wechsler&Tinker2018}.
We here adopt throughout the simplest assumption that quiescent and starforming galaxies share the same underlying SMHM relation, and note that of our core results do not qualitatively depend on this working assumption.

\subsection{Halo occupation and implied size function}
\label{sec:hod}
As a first step, in Figure \ref{fig:cartoon_hod} we show the SMHM relation and its scatter for the four toy models, as well as the distribution of the host halos (i.e. the halo occupation distribution) and the implied size functions. Figure \ref{fig:cartoon_hod} reveals that different SMHM relations and their scatter $\sigma_{\rm SMHM}$ provide significantly different size functions, that necessarily stem from distinct host halo occupation distributions. Thus, the size functions are completely regulated by the way the SMHM relation maps galaxies into haloes.  In particular, it is relevant to highlight the following features when comparing different models for the input SMHM relation:
\begin{itemize}
    \item \textbf{Model 1 vs Model 3.} A change in the high-mass slope of the SMHM relation generates an overall lower number density of MGs, but the mean of the halo occupation distributions and related size functions are fairly similar in the two cases.
    \item \textbf{Model 1 vs Model 2} and \textbf{Model 3 vs Model 4.}
    Even when the shape of the SMHM relation is identical, { if we allow for the scatter $\sigma_{\rm SMHM}$ to evolve with redshift, and in particular to increase at earlier epochs, then} the implied halo occupation distribution drastically changes compared to the case with constant scatter. In the former case, a higher proportion of small MGs are hosted in less massive haloes at higher redshift, and the mean halo occupation and galaxy size exhibit a stronger evolution, as quantitatively described below.
\end{itemize} 

\subsection{Implied size evolution}
\label{sect:size_evolution}

{\bl{S14}} showed that, on the assumption that $R_e\propto R_h$ at all epochs, the progressive increase in virial radii and in the number densities of massive dark matter haloes, were sufficient conditions to produce, when averaging over the full population, a strong size evolution in the sizes of massive galaxies.

\begin{figure*}
 
 \centering
 \includegraphics[width= \textwidth]{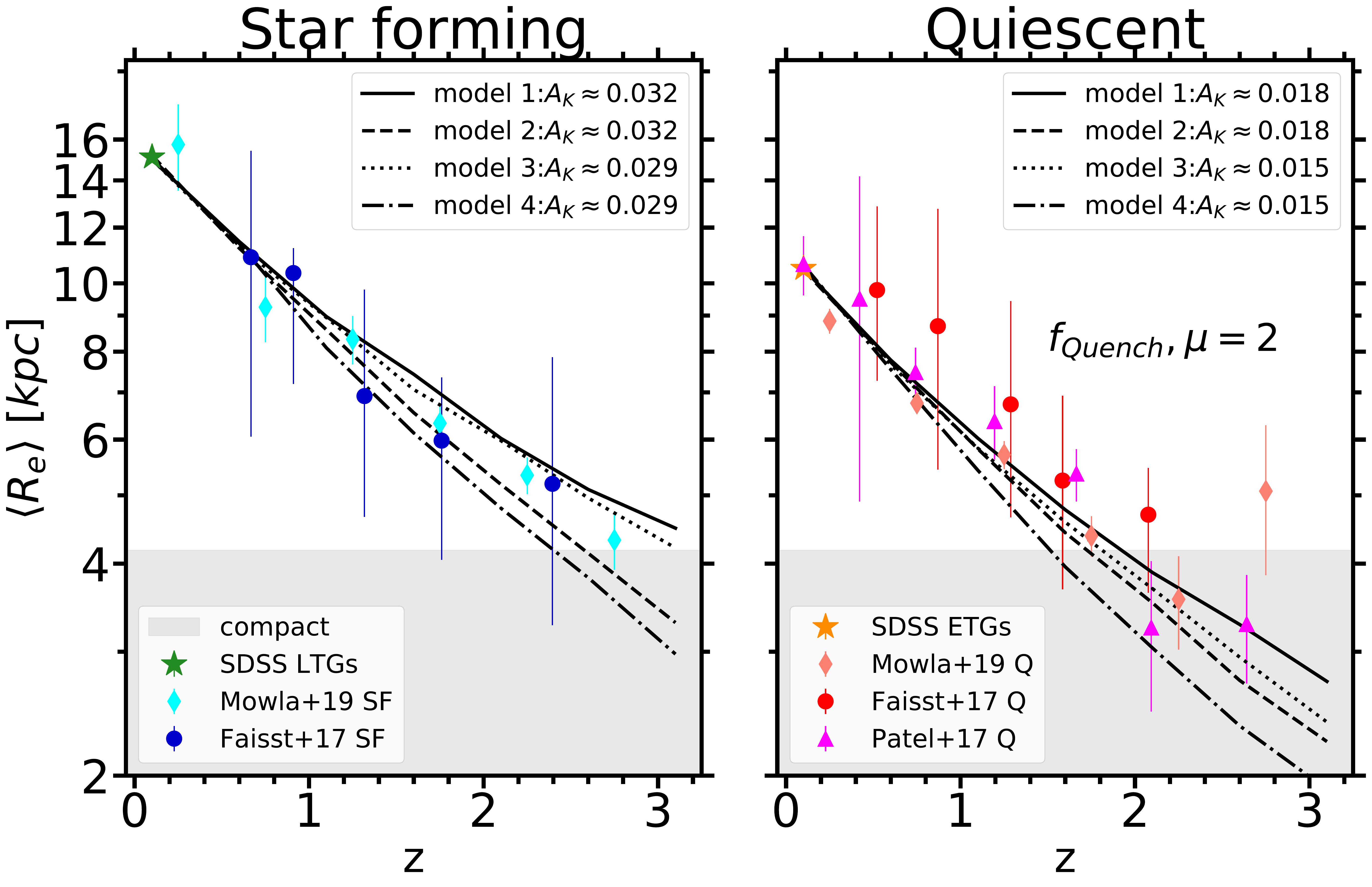}

\caption{The size evolution from the K13 model for starforming (left) and quiescent (right) MGs. The black lines indicate the four toy models outlined at the beginning of Section \ref{sec:systematic_analysis} and coupled with the $f_{Quench}$ model introduced in Section \ref{sec:fquench} with $\mu=3$, as an example. Data are the sizes of MGs from \citet{Mowla+18} (diamonds), \citet{Faisst+17} (circles), \citet{Patel+17} (triangles). We also add SDSS estimates for the sizes of Massive Late type and Early type galaxies (green and orange stars respectively) from \citealp{Zanisi+20}. The normalization $A_K$ in each panel is chosen to match SDSS observations. Notably, a constant normalization $A_K$ is able to reproduce observations. Moreover, $A_K$ is lower for shallower high-mass-end slopes of the SMHM (Models 1 and 2), while the opposite is true for steeper SMHM relations. This indicates that $A_K$ and $\delta$ are degenerate in our model.}
\label{fig:size_ev_SF_Q}
\end{figure*}

Figure \ref{fig:size_ev_SF_Q} confirms and further extends the claim by S14. By using, for each of our four toy models, a constant proportionality $R_e=A_K\times R_h$ calibrated at $z=0.1$ (see Appendix \ref{app:sloan}), as labelled, we are always able to reproduce the strong redshift evolution seen in the available data \citep[][]{Faisst+17,Patel+17,Mowla+18}, irrespective of the exact input SMHM relation. Models with an evolving $\sigma_{\rm SMHM}$ tend to predict up to less than 50\% faster size evolutions, well within the variance currently found in the data. We distinguish between starforming and quiescent galaxies via the $f_{Quench}$ model with $\mu=2$. Varying the $\mu$ parameter has little effect on our results, as it can be easily compensated by a relative variation in $A_K$ and/or in the SMHM relation. Indeed, the $A_K$ retrieved for starforming and quenched MGs appear to be systematically different and such difference persists even when adopting 
distinct SMHM relations as, for example, in \citet{Moster+18}, for which we find  $A_{K,SF}\approx0.023$ and $A_{K,Q}\approx 0.016$.

\subsection{Implied statistics of compact MGs}
\label{sec:fquench_compact}
\begin{figure*}
    \centering
    \includegraphics[width=0.8\textwidth]{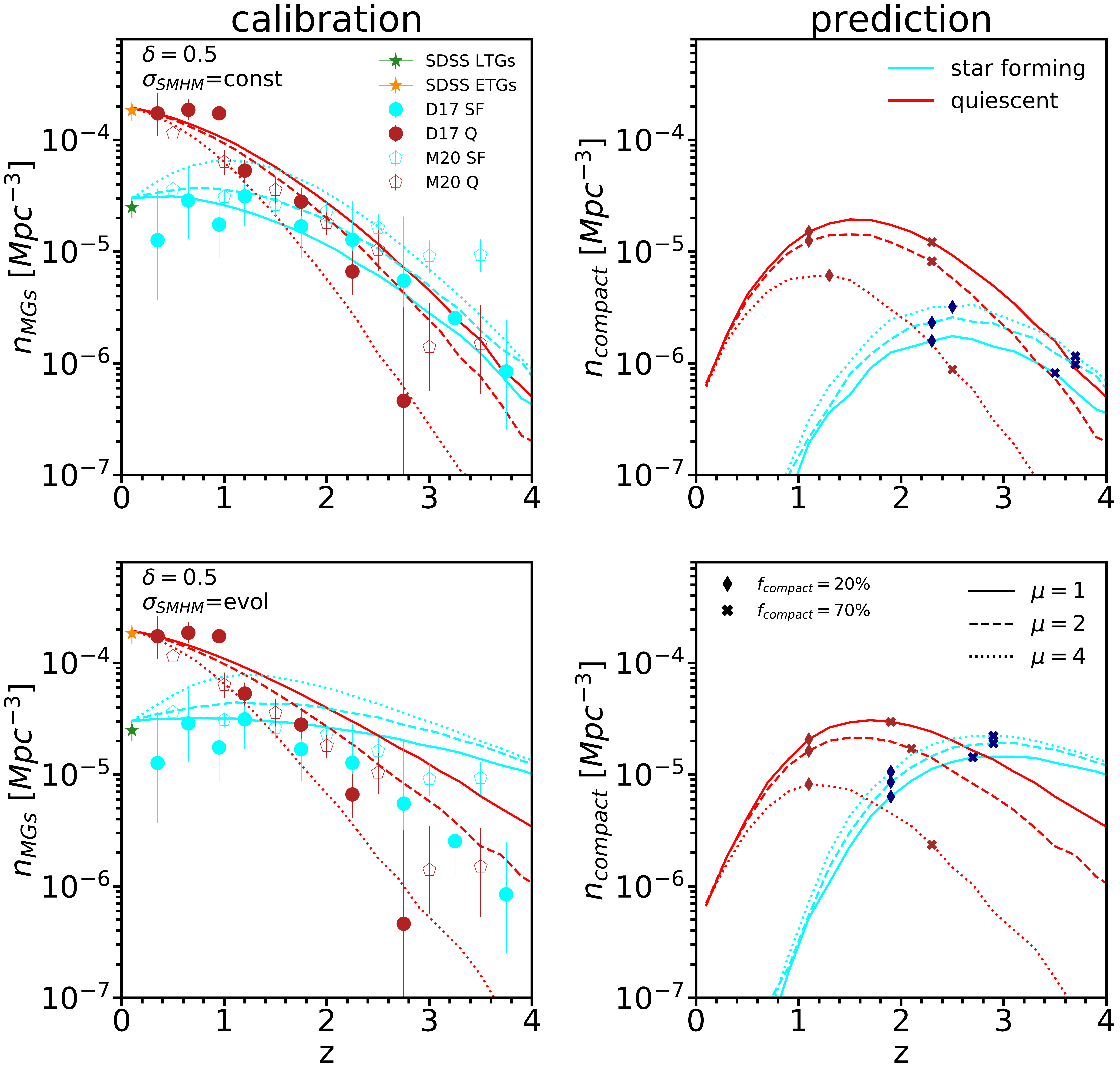}
    \caption{\emph{Left}: The number density of starforming and quenched MGs (cyan and red line respectively) for models 1 and 2. Solid, dashed and dotted lines are obtained adopting $\mu=1,2,4$ respectively. \emph{Right:} Prediction for the number density of compact MGs for the two models and the different values of $\mu$. Note that the fraction of compact MGs increases at early times. Filled diamonds and crosses indicate the time where 20\% and 70\% of the population of MGs (either starforming or quenched) are compact. The comparison data in the left column are from the SDSS 'PyMorph' photometry at z=0.1 \citep{Meert+15}, \citet{Davidzon+17} and \citet{McLeod+20} (in the wo latter cases the masses have been shifted by 0.15 dex to account for M/L differences with our SDSS data, see \citealt{Grylls+20_satellites}). Note that the data points were retrieved from the Schechter fits provided in the two studies, extrapolated in our mass range. With the caveat that different definitions of quiescence are adopted in observations, we note that Model 1 is favoured by current data if $\mu\approx 2-3$. Model 2 might provide a better fit to data if the number density of starforming MGs is underestimated at high redshift \citep{Franco+18_opticallyDark,Smail+21}.}
    \label{fig:cartoon_fquench}
\end{figure*}

In Figure \ref{fig:cartoon_hod} we showed that the shape and scatter of the SMHM have a significant impact on the number density of compact galaxies, a feature that was not investigated by previous studies. We explore these trends more quantitatively here for our toy models. The top and bottom panels of Figure \ref{fig:cartoon_fquench} show the predictions of Model 1 (constant scatter) and Model 2 (evolving scatter) for the number density of MGs (left panels) and for only compact MGs (right panels), separately for quiescent (red) and starforming (cyan) galaxies and for different values of the quenching parameter $\mu$, as labelled (the predictions for Models 2 and 3 are very similar and reported in Appendix~\ref{app:compact}).
All models predict a similarly sharp rise in the number density of compact quiescent MGs (red lines) up to $z\sim 1.5-2$ and a subsequent more or less fast drop depending on the exact value of $\mu$ adopted. All models also predict the abundances of starforming compact MGs (cyan lines) to peak around the same redshift $z\sim 2.5$ with a weak dependence on $\mu$ but a strong one on scatter: a larger $\sigma_{\rm SMHM}$ at early epochs can increase by up to a factor of ten the predicted number densities of starforming compact MGs (bottom right panel). In Appendix \ref{app:compactness} we show that adopting other definitions of compactness \citep[e.g.,][]{Barro+13, vanderwel+2014,Gargiulo+17} does not alter the main  qualitative trends of Figure \ref{fig:cartoon_fquench}. 

The evolution of $n_{compact}$ that we predict for compact quiescent MGs is in qualitative agreement with observations of compact galaxies in a lower mass range ($10.5<logM_{star}/M_\odot<11.5$, \citealt{Cassata+11,Cassata+13, vanderwel+2014, Barro+13}). However, at present, current observations provide rather uncertain constraints on $n_{MGs}$ at high redshift (see \citealt{Kawinwanichakij+20} for a detailed discussion of the systematics). In addition, a secure determination of the number density of, especially compact, MGs is hampered by the seizable but still unknown number of optically dark starforming galaxies at high redshift (e.g., \citealt{Franco+18_opticallyDark,Wang+19_Hdropout_Nature,Wenjia+20_opticallydark, Smail+21}). Nevertheless, the results presented in Figure \ref{fig:cartoon_fquench} provide clear predictive trends for the evolution of compact and large MGs that, when compared with data from the next generation of observing facilities, will set tight constraints on the quenching mechanisms ($\mu$ parameter) and on the level of progenitor bias in the size evolution of MGs.

\vspace{1em}

\section{Discussion}
\label{sec:discussion}

\subsection{Progenitor bias scenarios and continuity equation}
\label{sec:progbias}

\begin{figure*}
    \centering
    \includegraphics[width=0.8\textwidth]{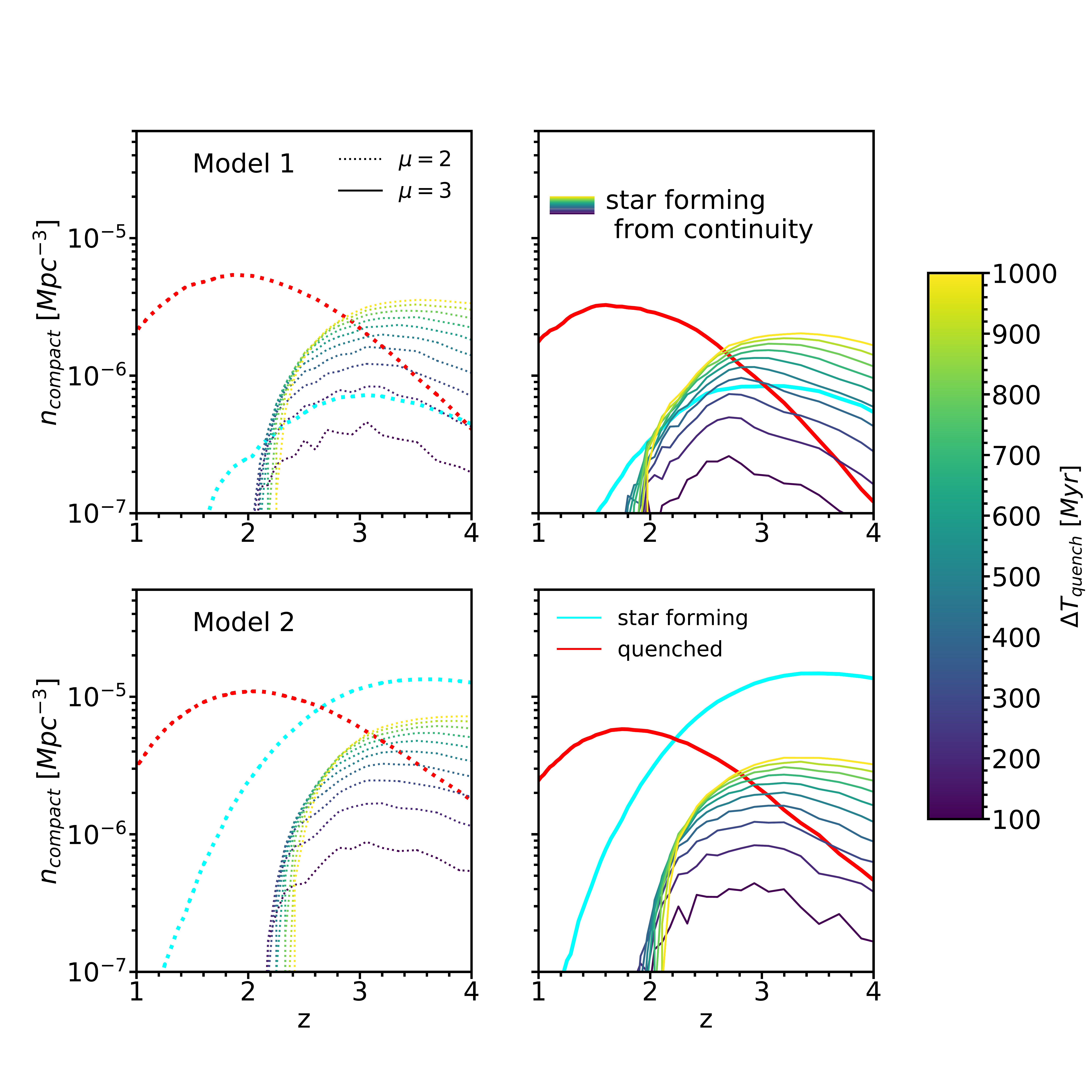}
    \caption{The number density of CSFMGs (cyan) and CQMGs (red) for Model 1 (top row) and Model 2 (bottom row). We adopt mu $\mu=2$ (dotted lines, left columns) and $\mu=3$ (solid lines, right column). The number density of compact starforming galaxies that would be obtained from continuity arguments (see eq. \ref{eq:continuity}) is shown for different values of the quenching timescale $\Delta T_{quench}$. Increasingly larger values of $\Delta T_{quench}$ are shown with increasing brightness. Model 2 disfavours a continuity scenario. In Model 1 continuity is broadly achieved if $\Delta T_{quench}\approx 200-400$ Myr for $\mu=2$ and $\mu=3$ respectively. Results for Models 3 and 4 can also be found in the \textbf{online supporting material}, and are qualitatively similar.}
    \label{fig:continuity}
\end{figure*}

We have demonstrated that all our models are able to produce a strong evolution in the average effective radius of the MG populations (Figure \ref{fig:size_ev_SF_Q}). On the other hand, Figures \ref{fig:cartoon_fquench} and \ref{fig:cartoon_fquench_2} clearly show that in all models $n_{compact}$ decreases below $z\sim1.5-2$. The peak of the abundance of compact quenched MGs corresponds to compact fractions of $\sim 20-40$\%. Thus, the ensuing disappearance of compact galaxies as the Universe ages strongly suggests that $\sim 20-40$\% of the quenched MGs that were present at $z\sim 1.5-2$ have grown in size \emph{individually} \citep[e.g.,][]{Trujillo+11_mergersSincez1,Carollo+13,vanderwel+2014,Fagioli+16,Faisst+17,Stockmann+20}. However, it is worth pointing out that this corresponds to only $\sim 10-15$\% of the quenched MGs that are present today (for the case of constant and evolving $\sigma_{\rm SMHM}$ respectively).

At $z\gtrsim 2$ all models instead predict a strong increase in the number density of compact MGs, suggesting that, in line with a number of observational studies \citep[e.g.,][]{Barro+13, Cassata+13}, a significant fraction of MGs form in a compact phase at early epochs, most probably due to gas dissipation following a merger \citep[e.g.,][]{Sparre&Springel2016} or an in-situ burst of star formation \citep[e.g.,][]{Lapi+11}.

An interesting question that has been discussed in the literature is whether compact quenched galaxies are the descendants of compact starforming galaxies \citep[e.g.,][]{vandokkum+15,Barro+17,GomezGuijarro+19_compacts_old_starbursts}. For example, based on number density conservation arguments, \citet{Barro+13} proposed that compact starforming galaxies with $10.5<logM_{star}/M_\odot<11.5$, passively evolve into quenched compact galaxies in a timescale of $\sim 800$Myr. Following \citet{Barro+13}, we here develop basic continuity equation models without mergers in which compact starforming MGs (CSFMGs) naturally evolve into compact quenched MGs (CQMGs) on a given timescale $\Delta T_{quench}$ as
\begin{equation}
\label{eq:continuity}
    n_{CSFMGs}(t) = n_{CQMGs}(t+\Delta T_{quench}) - n_{CQMGs}(t)  
\end{equation}
in which $\Delta T_{quench}$ is allowed to vary between 100 and 900 Myr, $t$ is the age of the Universe, and $n_{CQMGs}(t)$ and $n_{CSFMGs}(t)$ are the cumulative number densities of quiescent and starforming compact MGs above $M_{star}>10^{11.2}\, M_{\odot}$.
Figure \ref{fig:continuity} shows the results of applying Eq~\ref{eq:continuity} to the $n_{CSFMGs}$ extracted from Models 1 and 2 (see \textbf{online supporting material for Models 3 and 4)} with quenching parameters $\mu=2,3$ as a reference (the results derived for other values of $\mu$ are  included in the online supporting material.)

As reported in the bottom panel of Figure \ref{fig:continuity}, 
models characterised by a scatter $\sigma_{\rm SMHM}$ increasing at earlier epochs tend to disfavour a continuity scenario in which all CSFMGs gradually transition into CQMGs, as the number densities of CSFMGs (cyan lines) are always significantly larger than those of compact quiescent galaxies (red lines). Instead, models with a fixed $\sigma_{\rm SMHM}$ (Model 1, top row of Figure \ref{fig:continuity}) are broadly consistent with a progenitor-descendant scenario between CSFMGs and CQMGs for some choices of $\Delta T_{quench}$. In the specific, we find that $\Delta T_{quench}\approx200,300,400,900$ Myr for $\mu=2,2.5,3,4$ (data shown only for $\mu=2,3$, see \textbf{online supporting material} for $\mu=2.5,4$). \citet{Thomas+05} estimated an upper limit to the main star formation episode of local MGs around $\Delta T_{quench}\lesssim300$Myr (see their eq. 5), which would be consistent,  at face value,  with continuity in our constant $\sigma_{\rm SMHM}$ models with $2\lesssim\mu\lesssim 3$, in line with our preferred values of $\mu$ adopted in Figures \ref{fig:cartoon_fquench} and \ref{fig:cartoon_fquench_2}. We note that continuity arguments applied to Models 3 and 4 (see \textbf{online supporting material}) yield results that are qualitatively similar to Models 1 and 2 respectively.

\vspace{1em}

In a continuity scenario between CSFMGs and CQMGs (which can be produced by Models 1 and 3), little or no size evolution occurs during quenching. This conflicts with theoretical models where both size evolution and quenching occur almost simultaneously as a result of AGN activity, with a predicted expansion in size of a factor of $\gtrsim 2$ over very short timescales (i.e., $50-100$ Myr, \citealt{Ragone-Figueroa&Granato2011, Lapi+18_ETGs}). In other words, assuming a constant scatter $\sigma_{\rm SMHM}$ in the input SMHM relation, would be consistent with a two-stage formation scenario in which galaxies first quench and then grow via stochastic mergers \citep[e.g.,][]{Hopkins+09_spheroidScalingRelations_mergers, Oser+12}. Alternatively, an increasing $\sigma_{\rm SMHM}$ at earlier epochs would necessarily require within our framework that only a relatively minor fraction of the CSFMGs quench during their compact phase, a scenario more consistent with an AGN-driven size evolution. We note that an unbiased view of the size growth of MGs requires both optical-NIR observations as well as FIR-submm observations \citep[e.g.,][]{Barro+16,Tadaki+20,Sun+21}. Compact dust-enshrouded star formation activity can in fact occur over spatial scales a factor of $\sim$3 smaller that the $R_e$ measured in optical-NIR \citep[e.g.,][]{Puglisi+19,Jimenez-Andrade+19-compacts}. AGN activity in these galaxies might cause, along with quenching, a considerable evolution in size in a very short timescale \citep[e.g.,][]{Lapi+18_ETGs}. 

We conclude this Section by stressing the fact that our continuity models strictly apply to \emph{compact} MGs, which amount to a substantial fraction of the total population of quiescent MGs only at $z\gtrsim 2$ (see diamonds and crosses on Figure \ref{fig:cartoon_fquench}). It is evident from Figure \ref{fig:cartoon_fquench}, that all models predict an increase in the number density of the overall quiescent population at $z<2$ by up to an order of magnitude, a trend that cannot be driven by solely quenching of the starforming MGs as the number density of the latter is always significantly lower that those of quenched MGs at late epochs. Additional physical processes must be at play at $z<2$ in regulating the formation and sustenance of non-compact starforming MGs as well as the appearance of a large population of non-compact quenched MGs.

\subsection{The sizes of MGs as effective constraints to the galaxy-halo connection}
\label{sec:sizes_constraints_SMHM}
Providing firm constraints to the SMHM relation at different epochs can yield invaluable information on, e.g., the merger rates of MGs \citep{Grylls+20_pairfract}, the interplay between dark matter and baryonic physics \citep{Gu+16_scatterSMHM,Matthee+17}, the physical processes behind galaxy quenching \citep{Tinker+17_testingTheoriesWithScatter}.
Unfortunately, the shape and scatter of the SMHM relation are still highly debated (e.g.,\citealt{Bernardi+17_highmassend}).
In particular, there is a well-known degeneracy between the high-mass slope, $\delta$, and the dispersion, $\sigma_{\rm SMHM}$, of the SMHM relation (e.g. \citealt{Shankar+14_SMHM}). Similarly to \citet{Grylls+20_pairfract}, in the previous Sections we made use of toy models where only these two parameters are changed to explore their impact on the sizes of MGs. As shown above, SMHM relations with different values of $\delta$ and $\sigma_{\rm SMHM}$ result in distinct rates of size increase (Figures \ref{fig:size_ev_SF_Q}) and  number density evolution of compact MGs (Figures \ref{fig:cartoon_fquench} and Figure \ref{fig:cartoon_fquench_2}), which are ultimately a consequence of the different implied halo occupation distribution (Figure \ref{fig:cartoon_hod_introduction}). Our results therefore suggest that the $\delta-\sigma_{\rm SMHM}$ degeneracy may be broken by simultaneously fitting the size growth of MGs, the redshift evolution of the number density of compact MGs, and the number density evolution of the overall population of MGs, in other words by an accurate measurement of the full size function of MGs at different epochs, a goal that should be achievable with the aid of the next-generation observational facilities such as Euclid and LSST. We note that several previous semi-empirical studies aimed at probing the size evolution of galaxies (e.g.,  \citealt{Rodriguez-Puebla+17,Hearin+19,Behroozi+21}). However, they were are all limited by the use of only one SMHM relation and dispersion, which instead, if allowed to vary, can provide distinct structural evolutionary tracks for MGs.

It is important to highlight that the systematic uncertainties in measuring stellar masses and number densities of galaxies substantially affect the determination of the stellar mass function, and thus of the SMHM relation and size distributions of MGs at different epochs, possibly explaining at least part of the observational discrepancies in the numbers of compact galaxies reported in the literature (e.g., \citealt{Poggianti+13}).
\subsection{Covariance between the $R_e-R_h$ and the SMHM relations?}
\label{sec:covariance}

A further interesting issue that warrants more exploration is that of a possible covariance between the $R_e-R_h$ and the SMHM relation. In our framweork, the $R_e-R_h$ and the SMHM relations are closely intertwined. However, we did not consider an explicit correlation between the two relations which, instead, may be possible. For example, SDSS observations \citep{Bernardi+14} have shown that at fixed velocity dispersion (which is a proxy of halo mass, e.g. \citealt{Sohn+20_sigmacluster_sigmastar_mhalo}) brighter (i.e. more massive) galaxies have larger $R_e$ (\citealt{Shankar&Bernardi2009}). Such a trend may be captured by introducing a positive covariance between the scatters of the SMHM and of the K13 relation. We ran a few simple tests and found that adding this ingredient to our framework does not significantly affect the implied size evolution of MGs. 
When including a positive covariance, less massive galaxies tend to also naturally be the smallest galaxies. The covariance thus ultimately generates narrower size function at fixed stellar mass, where the abundance of compact galaxies is now only controlled by the dispersion in size at fixed halo radius $\sigma_K$. Therefore, a higher $\sigma_K$, we find, can produce the same amount of compact galaxies as in a model without covariance but with a proportionally lower value of $\sigma_K$. The degeneracy between $\sigma_K$ and a covariance between the SMHM and the K13 relations may be broken by probing the environmental dependence of galaxy size at fixed stellar mass, a task that is beyond the scope of the present work.

\subsection{Including concentration in the K13 model}
\label{sec:concentration}
Some authors have argued that galaxy sizes may be regulated also by halo concentration \citep{Jiang+19, Desmond2017_concentration, Desmond+18_SPARC}. Essentially, this "concentration model" is a modified version of the K13 model where an inverse proportionality between galaxy size and halo concentration is also considered,
\begin{align}
R_e  &= A_c \bigl (\frac{c}{10}  \bigr) ^{\gamma} R_h \nonumber \\ &= f(c)R_h
\label{eq:concmodel}
\end{align}
where we define $f(c)=A_c \bigl (\frac{c}{10}  \bigr) ^{\gamma}$ and $\gamma<0$. Here we will leave $\gamma$ as a redshift-independent free parameter. For the concentration, we adopt the concentration-mass relation by \citet{Dutton&Maccio2014},
\begin{equation}
\label{eq:concD14}
\log{c} = a+ b\log{M_h [M\odot]/10^{12}/h}
\end{equation}
with $a(z)=0.537 + (1.025-0.537)exp(-0.718z^{1.08})$ and $b(z) =-0.097 +0.024  z$. \citet{Dutton&Maccio2014} report a log-normal scatter of about $\sim 0.11$ dex which is independent on halo mass.

\begin{figure*}
    \centering
    \includegraphics[width=0.8 \textwidth]{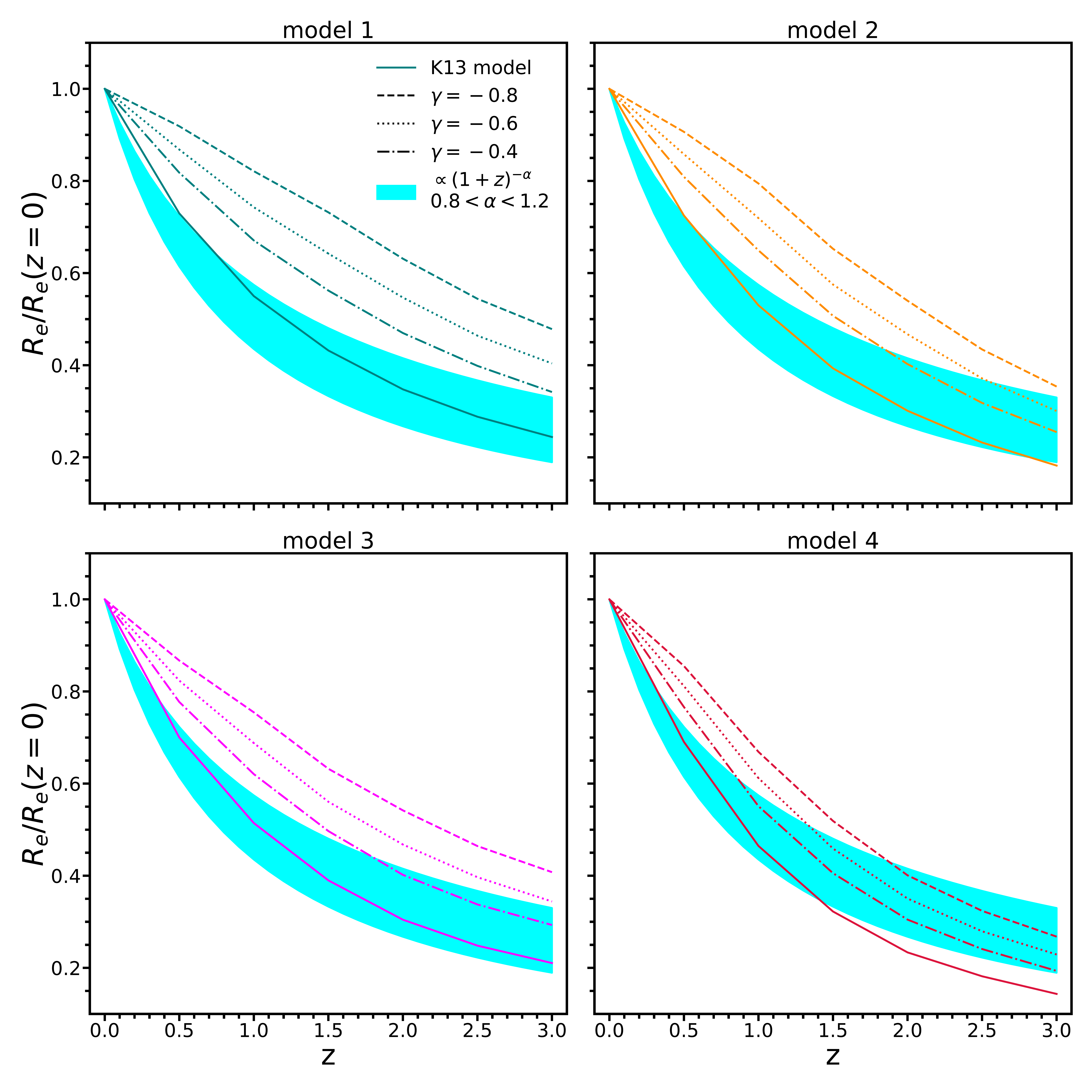}
    \caption{Size evolution inferred from the concentration model for $\gamma=-0.4,-0.6,-0.8$ (dot-dashed,dotted and dashed lines respectively). Top left is for Model 1, top right for Model 2, bottom left for Model 3 and bottom right for Model 4. In all panels, solid lines indicate the K13 model). The cyan shaded area broadly indicates the range of observational constraints allowed by current data ( $R_e \propto (1+z)^{-\alpha}$ with $-1.2<\alpha<-0.8$, see \citet{Faisst+17,Patel+17,Mowla+18}). All models struggle to reproduce the observed size evolution. Model 4, which has a shallow high-mass slope in the SMHM and for which an evolving $\sigma_{\rm SMHM}$ is implemented, provides a better match to the observed trend for some values of $\gamma$.}
    \label{fig:concentrationmodel}
\end{figure*}

\begin{figure}
    \centering
    \includegraphics[width=0.5 \textwidth]{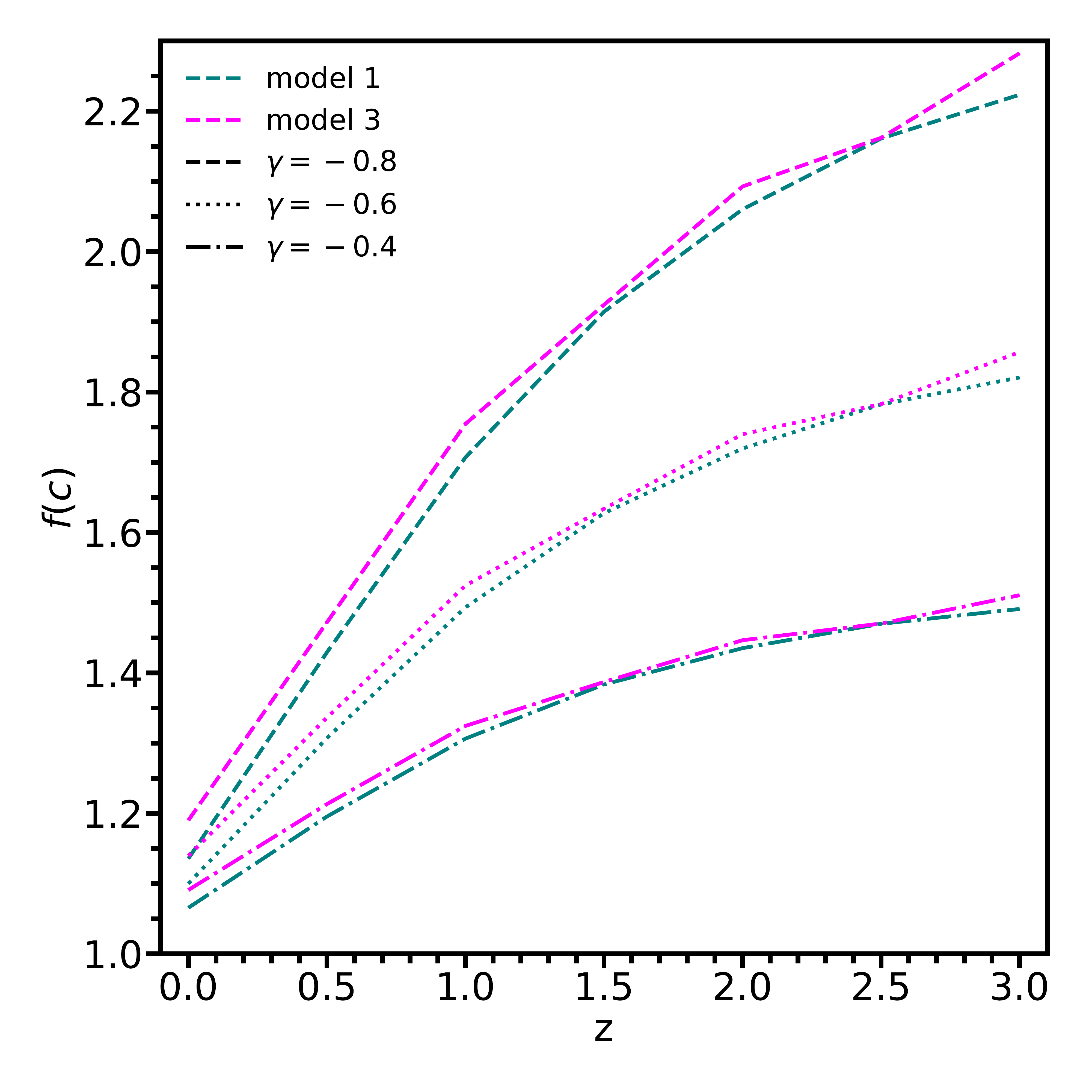}
    \caption{The redshift dependence of the factor $f(c)$ in the concentration model for $\gamma=-0.4,-0.6,-0.8$ (dot-dashed,dotted and dashed lines respectively), for Model 1 (teal lines) and Model 3 (magenta lines). $f(c)$ increases at earlier times, and it is weakly dependent on the SMHM relation.}
    \label{fig:fc}
\end{figure}

The results from the concentration model with $\gamma=-0.4,-0.6,-0.8$ are reported in Figure \ref{fig:concentrationmodel} for the four toy models explored in this paper (see Section \ref{sec:systematic_analysis}).
The most important feature of this Figure is that all models struggle to reproduce the size evolution of MGs, except for Model 4 characterised by a flat high-mass SMHM slope $\delta$ and an evolving scatter $\sigma_{\rm SMHM}$.
All models predict an increase in size at fixed stellar mass, with higher (absolute) values of $\gamma$ generating a shallower evolution. As $\gamma$ approaches zero, the trend tends to reduce to that of the K13 model, as expected. The departure from the K13 model is explained by the evolution of the factor $f(c)=\bigl(c/10 \bigr)^{\gamma}$ (equation \ref{eq:concmodel}, Figure \ref{fig:fc}), which has the effect of slowing down the evolution with respect to the K13 model. The predicted relatively slower size evolution in the concentration model is roughly independent of the input SMHM due the shallow correlation between halo mass and concentration (see eq. \ref{eq:concD14}). Although the concentration model struggles to reproduce a strong size evolution, as already noted by \citet{Jiang+19}, it cannot still be ruled out as current data may be underestimating galaxy sizes at high redshift due to surface brightness \citep[e.g.,][]{Ribeiro+16,Whitney+20} and/or colour gradients effects \citep[e.g.,][]{vanderwel+2014,Mosleh+17,Suess+19,Suess+20}.

\section{Conclusions}
\label{sec:conclusions}

In this work we developed accurate and transparent semi-empirical models to study the evolution of the size (effective radius $R_e$) function of massive galaxies (MGs, $M_{star}>10^{11.2}M_\odot$). We assumed an input stellar mass-halo mass (SMHM) relation to populate dark matter haloes with galaxies, and then assigned sizes to galaxies via an empirically-motivated linear and tight relation between $R_e$ and the host halo virial radius $R_{h}$. We varied the input SMHM relation to reflect the still substantial systematic uncertainties in the stellar mass function at both low and high redshift (see Footnote \ref{footnote:intro} in the Introduction). More specifically, we devised four toy models with different high-mass slopes and/or dispersions at fixed halo mass, $\sigma_{\rm SMHM}$, to probe their impact on the size function of MGs. In particular, we focused on the mean size growth and number density evolution since $z\sim 3$ of compact starforming and quiescent MGs, distinguished in the mocks via a simple halo mass-dependent quenching model with only one parameter \citep{Rodriguez-Puebla+15}. Our main results can be summarised as follows:

\begin{itemize}
    \item The shape and evolution of the size function is completely determined by the halo occupation distribution implied by each model. In particular, the number density of compact galaxies, $n_{compact}$, is a strong function of the scatter $\sigma_{\rm SMHM}$ (Figures \ref{fig:cartoon_hod_introduction} and \ref{fig:cartoon_hod}). 
    \item All models are able to broadly reproduce the fast size growth of starforming and quiescent MGs by simply assuming a redshift-independent $R_e-R_h$ relation with a different zero point for the two populations (Figure \ref{fig:size_ev_SF_Q}) and in ways largely independent of the shape of the input SMHM relation and of its scatter. 
    \item In all models, the number density of compact starforming MGs peaks at around $z\sim2.5$ and sharply declines at later times, while the peak in the number density of compact quiescent MGs is always delayed by a characteristic timescale which depends on the specific model (Figure~\ref{fig:cartoon_fquench}). Our findings thus suggest a size growth driven by newly formed MGs at $z\gtrsim1.5-2$, e.g. ``progenitor bias'', which plays a gradually lesser (but still important) role at $z\lesssim 1.5$. 
    \item In models in which the scatter $\sigma_{\rm SMHM}$ is strictly constant in time, we find that our predictions are consistent with a two-phase evolution scenario, in which compact starforming MGs first quench into compact quiescent MGs on a timescale of a few hundred Myr (Figure~\ref{fig:continuity}), and then grow in size (possibly via dry mergers). In models in which $\sigma_{\rm SMHM}$ is instead allowed to increase at earlier epochs, a significant proportion of quiescent MGs must increase their sizes before final quenching as in, e.g., AGN-driven size growth.
    \item We also implemented another variant of the models in which $R_e$ is proportional to virial radius via a halo concentration-dependent factor $f(c)$ \citep{Jiang+19}. We find that, at face value, this model struggles at reproducing the fast size growth of the population of MGs (Figure~\ref{fig:concentrationmodel}), although the data may be underestimating galaxy sizes at high redshift.
    
 \end{itemize}
 
All in all, our results support the view that an accurate measurement of the full size function of MGs, which will become available with the next generation of observing facilities such as EUCLID and the Nancy Grace Roman Space Telescope, will be able to set constraints on: i) the high-mass slope and scatter of the SMHM relation, ii) the rate of evolution of the number density of compact quiescent and starforming MGs  and the related degree of progenitor bias, iii) the quenching timescales of starforming MGs, and iv) the evolutionary processes (mergers versus AGN feedback) driving the structural evolution of MGs.

\section*{Acknowledgements}
We thank the anonymous referee for helping us improve the presentation of the results. L.Z. acknowledges funding from the Science and Technologies Facilities Council, which funded his PhD through the DISCnet Center for Doctoral Training. F.S. acknowledges partial support from a Leverhulme Trust Research Fellowship. We thank Ignacio Trujillo, Simona Mei and Adriana Gargiulo for useful comments and Nushkia Chamba for providing data tables for the measurements of $R_1$.

\section*{Data Availability}
All data will be shared upon reasonable request to the authors.
We provide our codes at \url{https://github.com/lorenzozanisi/SizeModels}.

\appendix
\section{The $f_{Quench}(z)$ relation}
\label{app:fquench}
\begin{figure}
    \centering
    \includegraphics[width=0.5 \textwidth]{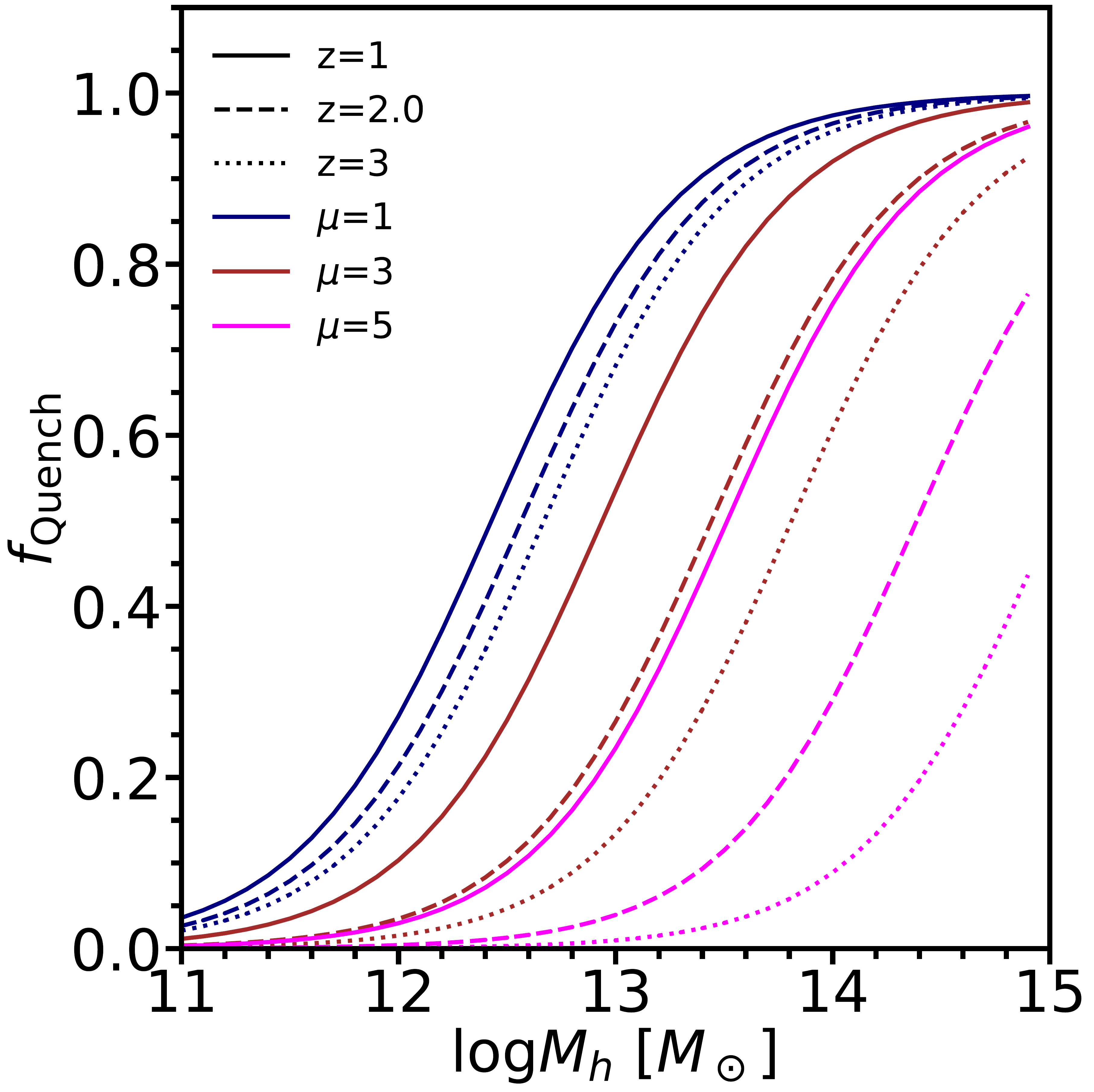}

    \caption{  The evolution of $f_{Quench}$ from eqs.  \ref{eq:fQevol} and \ref{eq:betaz} for $\mu=5$ (magenta lines) $\mu=3$ (red lines) and for $\mu=1$ (blue lines). In both cases, dotted lines, dashed lines and solid lines are for z=3,2,1 respectively. It can be seen that in models with a higher $\mu$ the halo mass scale above which galaxies are statistically quenched evolves much faster with redshift, and is higher at earlier cosmic times. }
    \label{fig:fquench_Mh_z}
\end{figure}

In Figure \ref{fig:fquench_Mh_z} we show the evolution of $f_{Quench}$ from eqs.  \ref{eq:fQevol} and \ref{eq:betaz} for $\mu=1,3,5$. It can be seen that in models with a higher $\mu$ the halo mass scale above which galaxies are statistically quenched evolves much faster with redshift, and is higher at earlier cosmic times.

\section{Calibration of $A_K$}
\label{app:sloan}
 We use Sloan Digital Sky Survey DR7 data \citep{Abazajian+09} to calibrate $A_{K,SF}$ and $A_{K,Q}$ at z$\sim$0.1. We create a mock catalog of MGs at z=0.1 as detailed in Section \ref{sec:methods}. We then constrain the normalization $A_K$ for starforming and quiescent galaxies by matching the mean size of the MGs in our mock catalog to the mean observed semi-major axis effective radius that best fits the light profiles of Massive Late Type Galaxies (LTGs) and Early Type Galaxies (ETGs) from the \citet{Meert+15, Meert+16, Dominguez-Sanchez+18} $r-$band photometric and morphological catalogues. We define LTGs as those objects for which $TType>0$ and ETGs those that have $TType\leq0$. We assume that all massive LTGs are starforming and all massive ETGs are quiescent. The light profile is truncated as in \citet{Fischer+17_truncation}.

\section{Definitions of compactness}
\label{app:compactness}

\begin{figure*}
    \centering
    \includegraphics[width=0.9\textwidth]{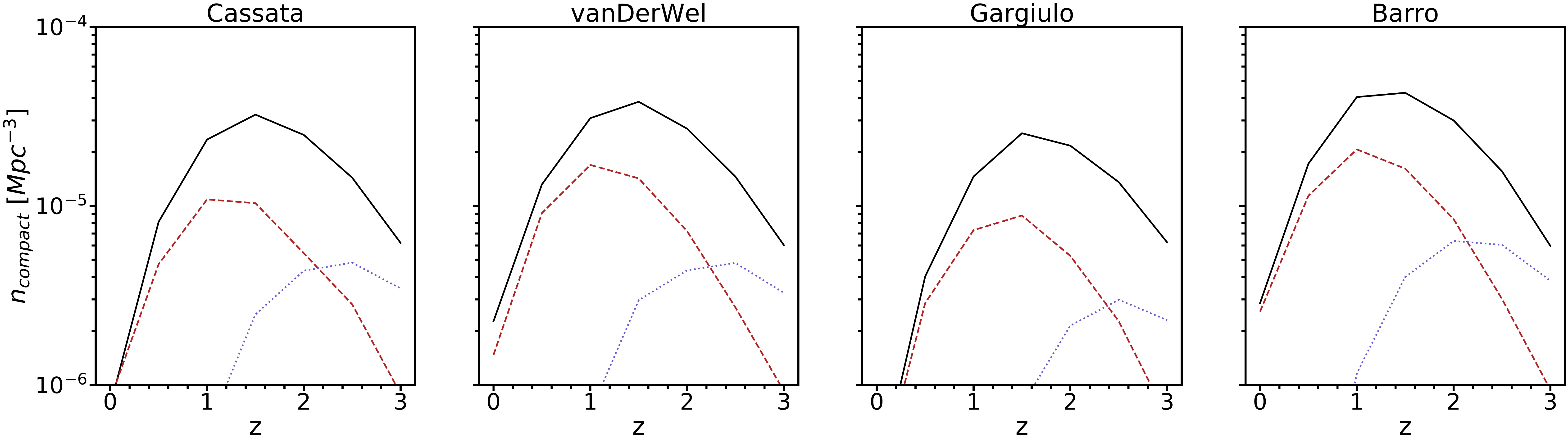}
    \caption{The evolution of the number density of compact quenched (red dashed lines) and compact starforming (blue dotted lines) MGs, for Model 1 and $\mu=3$, as in Figure \ref{fig:cartoon_fquench}. Cmpactness is defined as in \citet{Cassata+11} (left), \citet{vanderwel+2014} (center left), \citet{Gargiulo+17} (center right) and \citet{Barro+13} (right). Distinct definitions of compactness yield qualitatively very similar results, although quantitatively different.}
    \label{fig:supplementary_compacts}
\end{figure*}

In Figure \ref{fig:supplementary_compacts} we show the number density evolution of compact quenched and compact starforming MGs, for different definitions of compactness (including the one adopted in the main text of this paper, that is, that of \citealt{Cassata+11}), and for Model 1. 

\section{Compacts in models 3 and 4}
\label{app:compact}
\begin{figure*}
    \centering
    \includegraphics[width=0.8\textwidth]{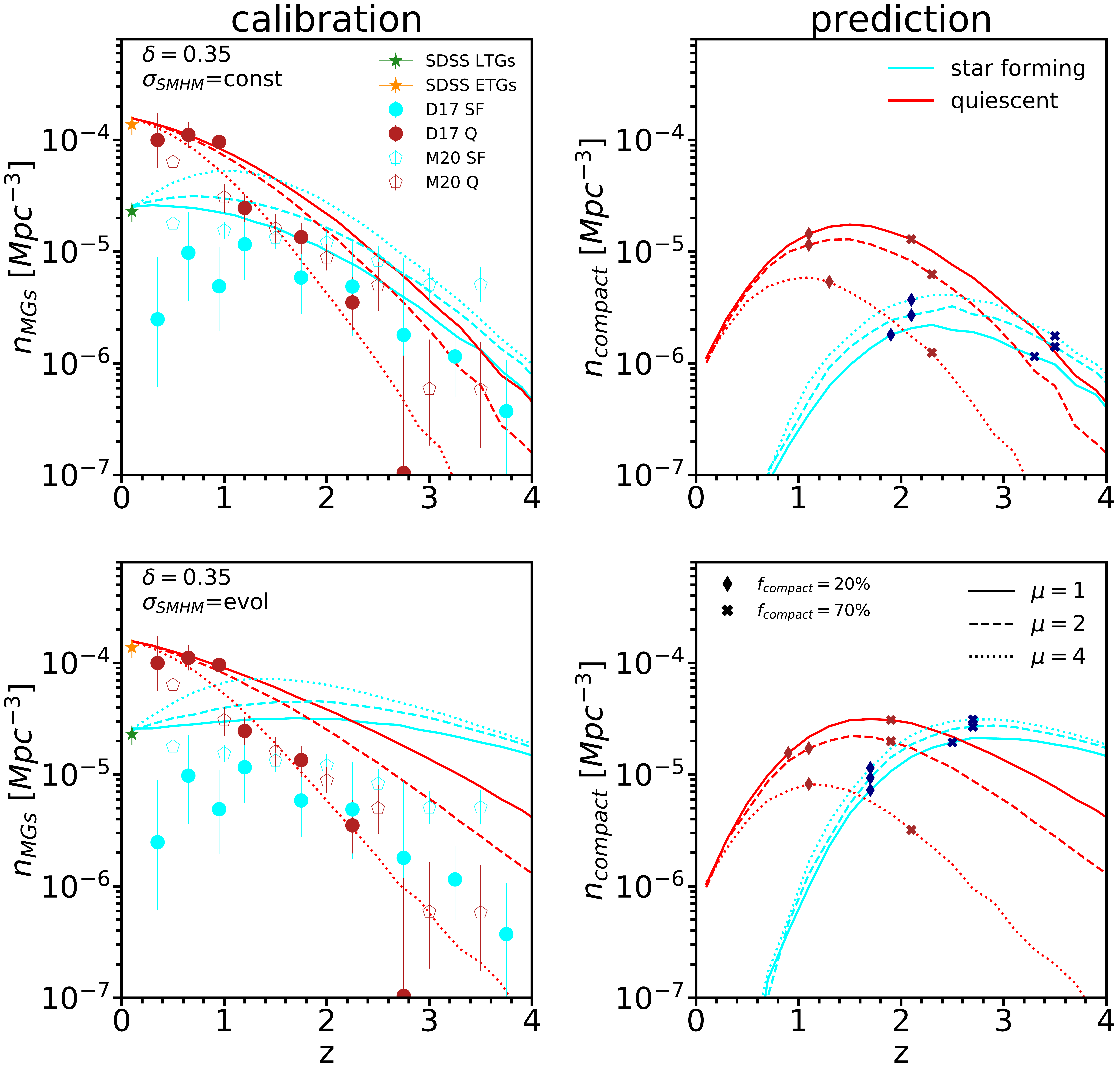}
    \caption{\emph{Left}: The number density of starforming and quenched MGs (cyan and red line respectively) for models 3 and 4. Solid, dashed and dotted lines are obtained adopting $\mu=1,2,4$ respectively. \emph{Right:} Prediction for the number density of compact MGs for the two models and the different values of $\mu$. The comparison data in the left column are from the SDSS 'cmodel' photometry at z=0.1, and \citet{Davidzon+17,McLeod+20} (not corrected for the M/L as it was done in Figure \ref{fig:cartoon_fquench}, see \citealt{Grylls+20_pairfract}) at higher redshift. }
    \label{fig:cartoon_fquench_2}
\end{figure*}

Figure \ref{fig:cartoon_fquench_2} shows the evolution of the number density of MGs and the corresponding predictions for the numbers of CQMGs and CSFMGs for Models 3 and 4. The results are qualitatively (but not quantitatively) similar to Models 1 and 2 (see Figure \ref{fig:cartoon_fquench}. In particular, most quenching models seem to disagree with the current data for starforming galaxies.

\section{Using other size estimators}
\label{sec:other_sizes}
Recent works have explored different definitions for the size of a galaxy. For instance,
 \citet{Mowla+19_SMHM_R80} and \citet{Miller+19_R80} put forward the idea that the radius that encloses 80\% of the light, $R_{80}$, might be more fundamental than $R_e$. This claim is based on the observation that (i) the size distributions of starforming and quenched galaxies are almost identical when using $R_{80}$ as opposed to the use of $R_e$, and (ii) that the shape and evolution of the $R_{80}-M_{\rm star}$ is reminiscent of the SMHM relation. Although we make explicit mention of $R_e$ throughout this paper, our model can be used to make predictions for the size evolution and the number density evolution of compact MGs, regardless of star formation activity, which may be interpreted in terms of $R_{80}$, rather than $R_e$. In particular, the K13 model would read
 \begin{equation}
     R_{80} = A_{K,80}R_h.
     \label{eq:K13_R80}
 \end{equation}
Using the values of $R_{80}$ for SDSS that we computed in {\bl{Z20}}, we find that that for MGs $\langle R_{80} \rangle \approx 4.2 \langle R_e \rangle$. Therefore, we expect $A_{K,80}\approx 4.2 A_K$.

\citet{Trujillo+20} and \citet{Chamba+20} used deep imaging of local galaxies to define $R_1$ as the radius that encloses the region within a physically-motivated mass surface density of $1 \rm M_\odot \ pc^{-2}$ (see also \citealt{SanchezAlmeida2020}). \citet{Trujillo+20} found that the scatter in the $R_1-M_{\rm star}$ relation is of the order of only $\approx0.06$ dex across five orders of magnitude, including the regime of MGs for which the relation, which is linear at lower masses, breaks. The $R_1-R_h$ relation would read
\begin{equation}
R_1=A_{K,1}R_h.
\label{eq:K13_R1}
\end{equation}
Using the publicly available catalog of $R_1$ measurements from \citet{Trujillo+20} and \citet{Chamba+20} we find that $\langle R_1 \rangle \approx7.8 \langle R_e \rangle$, which implies that the normalization of the K13 relation in eq. \ref{eq:K13_R1}, $A_{K,1}\approx 7.8A_K$. In Section \ref{sect:size_evolution} (Figure \ref{fig:size_ev_SF_Q}) we have shown that using a constant value of $A_K$ works remarkably well to describe the evolution of $R_e$. Whether this will be the case also for $R_1$ will be revealed by future deep high-redshift observations.

Lastly, we would like to highlight the fact that different size definitions provide different pieces of information: while $R_e$ is tight to the concentration of the light profile (see \citealt{Chamba+20}), $R_{80}$ probes also the outer regions of the galaxy. Likewise, $R_1$ has been proposed based on the gas mass density threshold required to initiate star formation. We believe that this does not necessarily make a definition of size more fundamental than another. Thus, it is possible that distinct definitions of galaxy sizes may be related to different physical processes generated by distinct galaxy-halo coevolution paths.

 \bibliography{papersize.bib}
\bibliographystyle{mnras}
\bsp	
\label{lastpage}

\end{document}